\pdfoutput=1 
\documentclass[prl,twocolumn,preprintnumbers,superscriptaddress,showpacs]{revtex4-1}

\usepackage{graphicx}
\usepackage{bm}
\usepackage{amsmath}
\usepackage{amssymb}
\usepackage[colorlinks,pdfstartview=FitH]{hyperref}
\hypersetup{linkcolor=blue,citecolor=blue,filecolor=black,urlcolor=blue}
\newcommand\sect[1]{\emph{#1.}---}

\def \be {\begin{equation} }
\def \ee {\end{equation}}

\def \bem {\begin{multline}}
\def \eem {\end{multline}}

\def \bes {\begin{subequations} }
\def \ees {\end{subequations}}

\def \pd {\partial}

\def \eq {Eq.~}

\def \c {\chi}

\def \e {\epsilon}

\def \k {\kappa}
\def \o {\omega}
\def \r {\rho}
\def \l {\lambda}
\def \m {\mu}
\def \n {\nu}
\def \s {\sigma}
\def \t {\tau}
\def \th {\theta}

\def \G {\Gamma}

\def \<{\langle}
\def \>{\rangle}
\def \+{\dagger}
\def \({\left(}
\def \){\right)}
\def \[{\left[}
\def \]{\right]}

\def \vk {\bm{k}}
\def \vj {\bm{j}}
\def \vB {\bm{B}}
\def \vx {\bm{x}}

\def \Ca {C^{(4)}_{1}}

\def \CS {\text{CS}}
\def \tr {\text{tr}}
\def \sph {\text{sph}}

\def \KK {\text{KK}}

\def \top {\text{Top}}

\def \anom {\text{anom}}
\def \norm {\text{norm}}

\def \ttheta{\theta}

\begin{document}

\preprint{RBRC-1099}
\author{Ioannis~Iatrakis}
\email{ioannis.iatrakis@stonybrook.edu}
\affiliation{Department of Physics and Astronomy, Stony Brook University, Stony Brook, New
York 11794-3800, USA}
\author{Shu~Lin}
\email{slin@quark.phy.bnl.gov}
\affiliation{RIKEN-BNL Research Center, Brookhaven National Laboratory,
Upton, New York 11973-5000, USA}
\author{Yi~Yin} 
\email{yyin@quark.phy.bnl.gov}
\affiliation{Physics Department, Brookhaven National Laboratory,
  Upton, New York 11973-5000, USA}

\title{Axial current generation by ${\cal P}$-odd domains in QCD matter}
\date{\today}

\begin{abstract}
The dynamics of topological domains which break parity (${\cal P}$) and
charge-parity (${\cal CP}$) symmetry of QCD are studied. We derive in
a general setting that those local domains will generate an axial
current and quantify the strength of the induced axial current. Our
findings are verified in a top-down holographic model.
The relation between the real time dynamics of those local domains and chiral magnetic effect is also elucidated. We finally argue that such an induced axial current would be phenomenologically important  in heavy-ion collisions experiment.
\end{abstract}
\maketitle 

\sect{Introduction}%
One remarkable and intriguing feature of non-Abelian gauge theories such as
the gluonic sector of quantum chromodynamics (QCD) 
is the existence of topologically non-trivial configurations of gauge
fields.
These configurations are associated with tunneling between different
states which are characterized by a topological winding number:
\be
Q_{W} =\int d^{4}x\, q\, ,
\qquad q=\frac{g^2\e^{\m\n\r\s}}{32\pi^2}\tr\(G_{\m\n}G_{\r\s}\)\,,
\label{}
\ee
 with $G_{\m\n}$ the color field strength.
While the amplitudes of transition between those topological states 
are exponentially suppressed at zero temperature,
such exponential suppression might disappear at
high temperature or high density\cite{Manton:1983nd,*Klinkhamer:1984di,*Kuzmin:1985mm,*Shaposhnikov:1987tw,*Arnold:1987mh,*Arnold:1987zg}. 
In particular,
for hot QCD matter created in the high energy heavy-ion collisions, 
there could be metastable domains occupied by such a topological gauge field
configuration 
which violates parity(${\cal P}$) and charge-parity (${\cal CP}$) locally.
We will refer to those topological domains as ``$\theta$ domain'' in
this letter (see also Refs.~\cite{Zhitnitsky:2011aa,*Zhitnitsky:2012im,*Zhitnitsky:2013hs} and
references therein for more discussion on the nature of ``$\theta$ domain'' ).

Due to its deep connection to the fundamental aspect of QCD, 
namely the nature of P and CP violation,
with far-reaching impacts on other branches of physics,
in particular cosmology,
the search for possible manifestation of those ``$\theta$ domains'' in
heavy-ion collisions has attracted much interest recently~\cite{Kharzeev:2007tn,Liao:2014ava} (see also \cite{Chao:2013qpa,*Yu:2014sla} for interesting effect of P and CP violation in related system).
A ``$\theta$ domain'' will generate chiral charge imbalance through
axial anomaly relation:
\be
\label{eq:anomaly}
\pd_{\mu}J^{\mu}_{A}= -2q\, . 
\ee
Furthermore,
the intriguing interplay between U(1) triangle anomaly (in
electro-magnetic sector) 
and chiral charge imbalance would lead to novel ${\cal P}$ and ${\cal CP}$ odd
effects which provide promising mechanisms for the experimental
detection of ``$\theta$ domains''.
For example,
a vector current and consequently the vector charge
separation will be induced in the presence of a magnetic field and chiral charge
imbalance.
Such an effect is referred as
the chiral magnetic effect (CME)
 \cite{Fukushima:2008xe}
(see Ref.~\cite{Kharzeev:2013ffa} for a recent review).
In terms of chiral
charge imbalance parametrized by the axial chemical potential
$\mu_{A}$,
CME current is given by:
$\vj_{V} = (N_c e\vB \mu_{A})/(2\pi^2)$.

\begin{figure}
  \centering
  \includegraphics[width=20em]{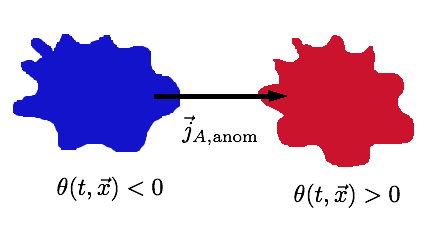}
  \caption{
\label{fig:ann}
(Color online) A schematic view of axial current due to
    the gradient of effective axion field $\ttheta(t,\vx)$ (c.f.~\eq\eqref{eq:j5new}).
Shaded areas illustrate ``$\theta$ domains'' (bubbles) with positive
$\theta$ (red)
and negative $\theta$ (blue).
The axial current flows from ``$\theta$ domains'' with smaller value
of $\theta$ to 
those with larger value. 
}
\end{figure}

To decipher the nature of ``$\theta$ domain'' through vector charge
separation effects such as CME,
it is essential to understand not only the distribution of
those chiral charge imbalance,
but their dynamical evolution as well. 
Previously,
most studies were based on
introducing chiral asymmetry by hand,
after which the equilibrium response to a magnetic field (or vorticity)
is investigated (see Ref.~\cite{Fukushima:2010vw} for 
the case in which the chirality is generated dynamically due to a particular color flux tube configuration).
In reality, such as in heavy-ion collisions experiment, 
however,
the chiral imbalance is dynamically generated through the presence
of ``$\theta$ domain''.
In this letter,
we study the axial current induced by inhomogeneity of ``$\theta$ domain'', 
which can be conveniently described by introducing a space-time dependent $\theta$ angle
$\theta(t,x)$ (c.f.~Refs.~\cite{Kharzeev:2007tn,Kharzeev:2007jp}).
One may interpret $\theta(t,\vx)$ as an effective axion field creating
a ``$\theta$ domain''. 
We show 
that the presence of $\theta(t,\vx)$ will not only generate
chiral charge imbalance,
it will also lead to an axial current (c.f.~ Fig.~\ref{fig:ann}):
\be
\label{eq:j5new}
\vj_{A} = \kappa_{CS}\bm{\nabla} \ttheta(t, \vx)\,  .
\ee
Such an axial current, 
to best of knowledge,
has not been considered in literature so far.

As it will be shown later,
our results are valid as far as the variation of $\theta(t,
  \vx)$ in space
is on the scale larger than $1/T$ (or mean free path of the system)
and the variation of $\theta(t,\vx)$ in time is on the scale longer
than the relaxation time of the system
but shorter than the life time of ``$\theta$ domain''.
It is therefore independent of the microscopic details of the system. 
While we are considering a system which is in the deconfined phase of QCD,
the resulting current bears a close resemblance to that in the superfluid.
One may interpret the gradient $\bm{\nabla} \theta(t,\vx)$ in \eq\eqref{eq:j5new}
as the ``velocity'' of ``$\theta$
domain'',
similar to the case of superfluid that the gradient of the phase
of the condensate is related to the superfluid velocity. 
Moreover, we will show that
the changing rate 
$\pd_{t}\theta(t,\vx)$ is related to the axial chemical potential appearing in the chiral
magnetic current 
again similarly to the ``Josephson-type equation'' in superfluid.
The relation between $\mu_{A}$ and $\pd_{t}\theta(t,\vx)$ is suggested
in Ref.~\cite{Fukushima:2008xe}.
We will show how such a connection is realized in a non-trivial way.
%
%

\sect{The axial current in the presence of $\theta(t,\vx)$}%
In this section, 
we will derive \eq\eqref{eq:j5new} and 
the constitute relation of $j^{\mu}_{A}$ in the presence of
$\theta(t,\vx)$.
The expectation value of $q$ induced by $\theta$
in Fourier space,
is given by
$q(\o, \vk)= -G^{qq}_{R}(\o, \vk)\theta(\o, \vk)$ 
where $G^{R}_{qq}(\o, k)\equiv -i\int d^4x
e^{-i\vk\cdot\vx+i\o t}\<\[q(t,\vx), q(0,0)\]\>\Theta(t)$ is the
retarded correlator of the density of topological charge density $q$.
For $\o, \vk\ll T$ (or inverse of the mean free path),
one may expand $G^{qq}_{R}(\o, k)$ up to $O(\o^2,\,k^2)$:
\be
\label{eq:GqqR_hydro}
G^{qq}_{R}(\o, k) = -\chi_\top+
\frac{1}{2}\[- i \frac{\Gamma_{\CS}}{T} \o - \kappa_{\CS} k^2 + \tau_{CS}\o^2\]\, , 
\ee
Here the first term is the topological susceptibility. It is highly
suppressed in de-confined phase, as indicated by both lattice
measurement and holographic calculation
\cite{Bonati:2013dza,Bergman:2006xn}. 
We will ignore $\chi_\top$ from below.
 $\Gamma_{\CS}$ in the second term is the Chern-Simons diffusion rate
and $\k_{\CS}$ and $\t_{\CS}$ are new transport coefficients.
Combining \eq\eqref{eq:GqqR_hydro} and the anomaly relation \eqref{eq:anomaly},
we have in real space :
\be
\label{eq:qtheta}
\pd_{\mu}j^{\mu}_{A}
=-2 q(t,\vec{x})
=
\(\frac{\Gamma_{\CS}}{T} \pd_{t} 
+\kappa_{\CS}\bm{\nabla}^2_{x}-
\tau_{\CS}\pd^2_{t}\)\ttheta(t,\vx)\, . 
\ee
To proceed, 
we divide $j^{\mu}_{A}$ into two parts:
$j^{\mu}_{A}=j^{\mu}_{A,\anom}+j^{\mu}_{A,\norm}$.
Here, 
we require $j^\mu_{A,\anom}$ to satisfy anomaly equation, 
i.e. , 
$\pd_{\mu}j^{\mu}_{A,\anom}=-2q$.
Consequently, 
the remaining part $j^{\mu}_{A,\norm}$ 
is conserved:
$\pd_{\mu}j^{\mu}_{A,\norm}=0$.
In general, 
the above division is not unique. 
However, 
if we further require that $j^{\mu}_{A,\anom}$ to be local in
$\ttheta$, i.e.
$n_{A, \anom}, \vj_{A,\anom}$ must be expressed in terms of $\theta(t,\vx)$ and
its gradients,
$j^{\mu}_{\anom}$ can then be determined uniquely from \eq\eqref{eq:qtheta} as
follows.
We start our analysis with $\vj_{A,\anom}$.
 By taking the static limit of \eq\eqref{eq:qtheta} and noting $\vj_{A,\anom}$
transforms as a vector under $SO(3)$ spatial rotation, 
one finds that $\vj_{A,\anom}$ have to be expressed in gradient of
$\ttheta$ with the magnitude fixed by \eq\eqref{eq:qtheta}:
\be
\label{eq:jx_anom}
\vj_{A,\anom} =
\kappa_{\CS}\bm{\nabla}\ttheta+ {\cal O}(\pd^2)\, ,
\ee
as was advertised earlier.
Similarly,
taking the homogeneous limit of \eq\eqref{eq:qtheta}
gives zeroth component of $j^{\mu}_{A,\anom}$:
\be
\label{eq:jt_anom1}
j^{t}_{A,\anom} =
 \frac{\G_{\CS}}{T}\ttheta 
-\tau_{\CS}\pd_{t}\ttheta
+ {\cal O}(\pd^{2})\, .
\ee

It is worth pointing out that $\k_{\CS}$ appearing in
\eq\eqref{eq:GqqR_hydro} is accessible by the lattice. To see that, we note in the static limit
\be
\label{eq:GF}
G^{R}_{qq}(\o=0,\vk)
= -\chi_{\top}-\frac{1}{2}\kappa_{\CS} k^2\, , 
\qquad
k=|\vk|\, .
\ee
It is related to the Euclidean correlator $G^{\rm E}_{qq}$ by $G^{R}_{qq}(\o=0, k)=-G^{\rm E}_{qq}(\o=0, k)$, which promises the possibility of measuring
$\k_{\CS}$ on the lattice through the following Kubo-formula:
\be
\label{eq:kappa_Kubo}
\k_{\CS}= 
 \lim_{k\to 0}\frac{d^2 }{d k^2} G^{\rm E}_{qq}(\o=0,k)\, .
\ee
At zero temperature, $\kappa_{\CS}$ would coincide with the so-called
``zero-momentum slope'' of topological correlation function and is of
phenomenological relevance in connection with the spin content of the
proton (see
Ref.~\cite{Vicari:2008jw} and reference therein).
However, 
the importance of $\k_{\CS}$ in de-confined phase of QCD ,
to best of our knowledge,
has not yet been appreciated.
While $\chi_{\top}$ is highly suppressed in de-confined
phase, 
there is no reason for the suppression of $\k_{\CS}$.
Eq.~\eqref{eq:jx_anom} gives an explicit example where $\k_\CS$ is phenomenologically relevant.

\sect{Chiral charge imbalance, axial chemical potential $\mu_{A}$ and
  the real time dynamics of $\ttheta$}%
We are now ready to quantify the chiral charge imbalance due to the presence of $\theta(t,x)$.
We concentrate on the first term on the R.H.S. of \eq\eqref{eq:jt_anom1} and define
axial density generated by $\ttheta(t,\vx)$ as:
\be
\label{eq:jt_anom}
n_{A,\anom}(t,\vx)\equiv
j^{t}_{A,\anom}(t,\vx) =
 \frac{\G_{\CS}}{T}\ttheta(t,\vx)
+ {\cal O}(\pd)\, .
\ee
\eq\eqref{eq:jt_anom} implies that a local ``$\theta$ domain(bubble)''
will induce a local axial charge density.
Further insight can be obtained by looking at the axial chemical
potential $\mu_{A}$ corresponding to $n_{A,\anom}$ in \eq\eqref{eq:jt_anom}.
Using the linearized equation of state $n_{A}=\chi \mu_{A}$ where
$\chi$ is the susceptibility,
we have:
\be
\label{eq:mu5_theta}
\mu_{A} = \(\frac{\Gamma_{\CS}}{\chi T}\)\ttheta
= \frac{\ttheta}{2\tau_{\sph}}
\, ,
\ee
where we have introduced the sphaleron damping rate $\tau_{\sph}$,
which can be related to the Chern-Simon diffusion rate $\Gamma_{\CS}$
by the standard fluctuation-dissipation analysis \cite{Rubakov:1996vz}
(see also \cite{Iatrakis:2015fma}):
 $\tau_{\sph}\equiv (2\chi T)/\Gamma_{\CS}$.
\eq\eqref{eq:mu5_theta} relating $\mu_{A}$ and $\ttheta$ is \textit{new} in
literature.
It can be connected to the argument of Ref.~\cite{Fukushima:2008xe} in
which $\mu_A$ is identified with $\pd_{t}\ttheta$.
\eq\eqref{eq:mu5_theta} implies that due to dynamical effects,
one should replace $\pd_{t}$ in the identification $\mu\sim
\pd_{t}\theta$ with $1/\tau_{\sph}$,
the characteristic time scale of sphaleron damping.
The above analysis suggests that while
relation Eqs.~\eqref{eq:jt_anom},\eqref{eq:mu5_theta} have already captured the real time dynamics of the
effective axion field $\theta(t,\vx)$,
namely  the sphaleron damping.

Finally,
let us briefly comment on the conserved part of the axial current
$j^{\mu}_{A,\norm}$.
Due to diffusion,
we expect from \eq\eqref{eq:jt_anom} that:
\be
\label{eq:j5_norm} 
\vj_{A,\norm} 
=-D\bm{\nabla}{n_{A,\anom}}
=-D \frac{\Gamma_{\CS}}{T}\bm{\bm{\nabla}}\ttheta\,  .
\ee
The conservation of normal part determines the time component as $j^t_{A,\norm}=-\int dt \bm{\nabla}\vj_{A,\norm}$. It depends on the history of normal part current, thus non-local in $\theta$. It is also higher order compared to $j^t_{A,\anom}$.
For positive $\k_{\CS}$,
axial current induced by ``$\theta$ domain'' \eqref{eq:j5new} is opposite to the diffusive
current \eqref{eq:j5_norm}.
We now argue that $\k_{\CS}$ is always positive by noting that a
non-zero $\theta$ will shift the action of the system by
$S_{\theta}= \int d^4x q\theta$.
Using the expression for $q$ in \eq\eqref{eq:GF},
one finds that in the static limit,  $S_{\theta}= -(k_{\CS}/2)\int d^4x
(\bm{\nabla} \theta)^2$.
Therefore $\k_{\CS}$ might be interpreted as coefficient of kinetic term of ``axion field'' $\theta$ and must be positive \footnote{We stress that our $\theta$ field originates from topological fluctuation of gluons. It should not be confused with $\theta$ field from fermionic quasi-zero mode in \cite{Kalaydzhyan:2012ut}}.

\sect{The holographic model}%
The discussion above does not rely on the microscopic details of the
theory.
We would like to confirm our findings in a top-down holographic model, 
namely,  Sakai-Sugimoto model \cite{Sakai:2004cn,Sakai:2005yt},
which at low enegy is dual to the four-dimensional $SU(N_c)$ Yang-Mills with massless quarks in large $N_c$ and strong coupling.
The deconfined phase of the field theory is dual to  the $D4$ black-brane metric, which is a warped product of a $5d$ black hole 
 and $S^{1}\times S^{4}$, \cite{Witten:1998zw,Aharony:2006da}. 
For the present work, we will consider field fluctuations with trivial
dependence on $S^{1}\times S^{4}$, thus we only need the $5d$ black hole part of the metric:
\begin{equation}
\label{metric}
ds^2=\(\frac{u}{R}\)^{3 \over 2}\(-f(u)dt^2+d{\vec x}^2  \)
+\(\frac{R}{u}\)^{3 \over 2}\frac{du^2}{f(u)} \, ,
\end{equation}
where $f(u)=1-(u_H/u)^3$ and $u$ is the holographic coordinate with $u=\infty$ the boundary and
$u=u_{H}$ the horizon. 
 $u_{H}$ are related to the temperature of the system by 
$4\pi T = 3\sqrt{u_{H}/R^3}$. The flavor degrees of freedom are introduced by a pair of $D8/{\bar D}8$ probe branes, separated
 along the $S^{1}$ direction, \cite{Sakai:2004cn}. The probe branes do not back-react on the geometry.
 
We will compute axial density $n_{A}$ and axial current $j_{A}$ along one
particular spatial direction, say ``x''
direction in the presence of a source, $\theta(t,x)$.
To this end, we consider excitation of axial gauge field $A_M$ of the
$D8\/{\bar D}8$ branes, with its field strength
$F_{MN}=\pd_MA_N-\pd_NA_M$ and Ramond-Ramond $C_1$ form. 
The index $M$ runs over $t,x,u$ and the
rest of the components can be consistently set to zero. The source
$\th(t,x)$ is related to $C_1^{(4)}$, the component of $C_1$
along $S_1$ by $2\pi R_4C_1^{(4)}=\th$, where
$R_4$ is the radius of $S_1$. 
Following the holographic correspondence,
the axial current $j^{\mu}_{A}$ is dual to 
the axial gauge field, $A_{M}$ and the topological charge density
$q$ is dual to $C_1^{(4)}$.
In the presence of $A_{M}$, we consider instead
components of Ramond-Ramond $C_{7}$
form(c.f.~Ref.~\cite{Sakai:2004cn}) $B_{M}$. The field strength of $B_M$,
 $G_{MN}=\pd_{M}B_{N}-\pd_{N}B_{M}$,
is related to combination of $A_{M}, C^{(4)}_{1}$ by:
$({\cal N}_G)/(uK)\epsilon^{LMN}\(2\pi R_4\pd_{L}\Ca + 2A_{L}\)=G^{MN}$
\footnote{The Levi-Civita symbol is fixed by $\epsilon^{txu}=1$.} 
by Hodge duality between $C_7$ form and $C_1$ form. Here $K=4\pi/3$ 
and ${\cal N}_G=(729\pi K^3u_H^2)/(4\l^3T^4R_4^2)$ with $\l$ the 't Hooft coupling.

After integrating over $S^{1}\times S^{4}$ and noting fields depend
only on $t,x,u$, we obtain the effective action,
which contains kinetic terms of
$F_{MN},G_{MN}$ 
and Wess-Zumino coupling between $F_{MN}$ and $B_{M}$\cite{Iatrakis:2015fma}:
\begin{multline}
\label{eq:holo_action}
S=\int d^4x du\,
\frac{1}{4}\bigg( -{\cal N}_{F}u^{5/2} F^{MN}F_{MN}-\frac{{\cal N}_{G}}{u}G^{MN}G_{MN} \\
-4K \epsilon^{LMN}B_LF_{MN} \bigg)\, ,
\end{multline}
In  action~\eqref{eq:holo_action}, 
${\cal N}_{F}=(8N_c\l^2T^3R_4)/(81u_H^3)$ 
The indices in \eq\eqref{eq:holo_action} are raised by 5d black hole
part of the full metric. The equations of motion following
from~\eqref{eq:holo_action} are given by
\begin{align}\label{eoms}
&\pd_M\(G^{MN}/u\)=K/({\cal N}_G)\e^{NPQ}F_{PQ}, \nonumber\\
&\pd_M\(u^{5/2}F^{MN}\)=K/({\cal N}_F)\e^{NPQ}G_{PQ}.
\end{align}

According to holographic correspondence,
the one point functions $n_{A}, j_{A}$ are given by the functional
derivative of the gravity on-shell action with respect to the boundary
values of $A_{t},A_{x}$. Using~\eqref{eoms}, we can then express
$n_{A}, j_{A}$ in terms of $G_{tx},F_{tx}$:
\begin{align}
\label{eq:JA_rep}
&n_A=\frac{2Ki\o G_{tx}-ik \({\cal N}_F u^{5/2} f\pd_uF_{tx}\)}{\o^2-k^2f}\big
  |_{u\to\infty}\, , \nonumber\\
&j_A=\frac{2Kikf G_{tx}-i\o \({\cal N}_F u^{5/2} f\pd_uF_{tx}\)}{\o^2-k^2f}\big |_{u\to\infty}\, . 
\end{align} 
We now need to solve the bulk equation of motion for $G_{tx}$ and
$F_{tx}$(see Eq.~\eqref{eq:G_eom} and Eq.~\eqref{eq:F_eom} below) with appropriate boundary condition.
We impose the infalling wave condition at the black hole horizon. 
On the boundary, $G_{tx}$ has the following asymptotic expansion
\begin{align}\label{Gtx_exp}
G_{tx}=\frac{K}{2{\cal N}_G}(\o^2-k^2)\th(\o,k) u^2+\cdots-\frac{q(\o,k)}{K}+\cdots.
\end{align}
The $u^2$ term is proportional to $\th$ and the constant term gives
$q$. 
One could verify that \eqref{eq:JA_rep} and \eqref{Gtx_exp} indeed
reproduce the anomaly equation:
$\pd_tn_A+\pd_xj_A=2KG_{tx}(u\to\infty)=-2q$. 
We only keep the  constant term in near boundary expansion of $G_{tx}$ in the
  limit. 
The divergent terms should be removed by holographic
  renormalization procedure: e.g. the $\o^2-k^2$ factor in the leading
  $u^2$ term,
which is completely determined by the near boundary behavior of bulk
equation of motion, 
 indicates that it is a contact term that can be subtracted by a boundary counter term. In case of non-conformal backgrounds, as the Witten-Sakai-Sugimoto bulk space-time, the holographic renormalization procedure is carefully described in \cite{Kanitscheider:2008kd}. 
On the other hand, $F_{tx}$ is not sourced on the boundary, thus 
we set $F_{tx}(u\to\infty)=0$. 
Note that $K/{\cal N}_F\sim O(1/N_c), K/{\cal N}_G\sim O(1)$,
The back-reaction of $F_{tx}$ to $G_{tx}$ is $1/N_c$ suppressed.
Keeping leading contribution in $N_c$, we find the following equations
of motion for $G_{tx}$ and $F_{tx}$ from action \eqref{eq:holo_action}:
\be
\label{eq:G_eom}
\[\pd_{u}\(\frac{f}{u(\o^2-k^2
  f)}\pd_{u}\)-\frac{R^3}{u^4f}\]G_{tx}=0,
\ee
\begin{multline}\label{eq:F_eom}
\[\pd_{u}\(\frac{u^{5/2}f}{\o^2-k^2 f}\pd_{u}\)-\frac{R^3}{u^{1/2}f}\]F_{tx}=\\
2K\frac{k}{\o}\pd_{u}\(\frac{f}{\o^2-k^2f}\)G_{tx}\, .
\end{multline}

\sect{Results of holographic calculation}
We are interested in the solutions to Eq.~\eqref{eq:G_eom} and Eq.~\eqref{eq:F_eom} in hydrodynamic regime,
i.e., $\o, k \ll 1/T$.
They can be found analytically,
order by order in $(\o/T, k/T)$, following standard
procedure in literature (c.f.~Refs.~\cite{Kovtun:2005ev,Iqbal:2008by}).
The full expressions and details of the calculations are straightforward but lengthy
and will be reported in a forthcoming paper
\cite{Iatrakis:2015fma}.
In order to compute $n_{A},j_{A}$, we only need their
near-boundary expansions:
\begin{multline}
\label{eq:G_sol}
G_{tx}=
\frac{K}{2{\cal N}_{G}}(\o^2-k^2)\theta u^2+
\frac{K u^2_{H}\th}{{\cal N}_{G}}\big[
-i \o \(\frac{u_H}{R^3}\)^{1/2} \\
+\frac{1}{2}\(\o^2-k^2\)-c_0\o^2 \big]\, ,
\end{multline}
\be\label{eq:F_sol}
F_{tx}=-\frac{4K^2u_H^2k\theta}{3{\cal N}_{G}{\cal N}_{F}u^{3/2}}
\big[
-i\(\frac{u_H}{R^3}\)^{1/2}+\frac{2(\o^2-k^2)}{3\o}-c_0\o
\big]
\, .
\ee
where $c_0=(\sqrt{3}\pi+3\ln3)/18$.
From \eq\eqref{eq:G_sol},
we immediately read $q$ by using \eq\eqref{Gtx_exp}.
Further comparison with Eq.~\eqref{eq:qtheta} gives $\G_{\CS},
\k_{\CS}$ in Sakai-Sugimoto model\footnote{
The value of CS diffusion rate $\G_\CS$ has already been calculated in
Ref.~\cite{Craps:2012hd}.
Our value is four times the value computed in
  Ref.~\cite{Craps:2012hd} due to a different normalization.
}:
\begin{align}
\label{eq:CS_value}
\G_\CS=
\frac{2u^{2}_{H}K^3T^2}{{\cal N}_{G}}
=\frac{8\l^3T^6}{729\pi M_\KK^2}\, ,
\qquad
\k_\CS=\frac{3\G_\CS}{8\pi T^2}\, ,
\end{align}
%
where $M_\KK=1/R_4$ is the mas gap of the theory.
Now plugging Eq.~\eqref{eq:G_sol} and Eq.~\eqref{eq:F_sol} into Eqs.~\eqref{eq:JA_rep}, we recover the time component of axial current in Eq.~\eqref{eq:jt_anom} and spatial component as a sum of Eq.~\eqref{eq:jx_anom} and \eqref{eq:j5_norm}:
\begin{align}\label{j5_sum}
n_A=\frac{\G_\CS}{T} \theta 
\qquad
j_A=-ik\(D\frac{\G_\CS}{T}-\k_\CS\)\theta\, ,
\end{align}
where the diffusion constant $D=1/(2\pi T)$ in Sakai-Sugimoto model ~\cite{Yee:2009vw}.
%
%

\sect{Phenomenological implication in heavy-ion collisions}
In this letter,
we found a new mechanism for generating axial current
\eqref{eq:j5new} due to the inhomogeneity of  effective ``$\theta$
domains''.
We now estimate its magnitude in a hot QGP and examine its
phenomenological importance in heavy-ion collisions. 
We start by relating $\theta$ to $\mu_{A}$ using
\eq\eqref{eq:mu5_theta}.
In terms of
$L_{\theta}$,
the characteristic size of a ``$\theta$ domains'',
Eq.~\eqref{eq:j5new} can be then estimated as:
\be
\label{eq:jA_estimate}
j_{A,\theta}\sim \(\mu_{A}\k_{\CS}\)\(\frac{\tau_{\sph}}{L_{\theta}}\)
\sim \(\mu_{A}T^2\)\( \frac{\tau_{\sph}}{L_{\theta}}\)\, ,
\ee
where in the last step we have taken our holographic results
\eqref{eq:CS_value} which implies $\k_{\CS}\sim T^2$ as a crude estimate of $\k_{\CS}$ in QCD plasma.

We now compare \eq\eqref{eq:jA_estimate}  to axial current from other sources.
For QGP in the presence of magnetic field,
axial current can be generated by chiral charge separation effects(CCSE) \cite{Son:2004tq,*Metlitski:2005pr}.
Similar to CME,
the CCSE current is given by $\vj_{A,\text{CCSE}}=(N_c\m_Ve\bm{B})/(2\pi^2)$.
In heavy-ion collisions at top RHIC energy, 
$eB$ at early stage is of a few $m^2_{\pi}$ and consequently
$N_{c}e^2B/2\pi^2$ is at most the same order as $T^2$.
Moreover, 
in those collisions, most of $\mu_{V}$(or $\mu_{B}$) is generated from
fluctuations and is expected to be the same order as $\mu_{A}$.
We therefore conclude that axial current is at least comparable to
CCSE current if $\tau_{\sph}/L_{\theta}\sim {\cal O}(1)$ but could be
larger if $L_{\theta}<\tau_{\sph}$.
A similar argument also applies to the comparison to chiral electric
separation effect \cite{Huang:2013iia}.

The axial current \eqref{eq:j5new} studied in this work is induced by
topological fluctuation. In plasma with chiral charge, 
axial charge can also be generated by thermal fluctuation, which is
non-topological. Axial current can also exist as diffusion of
such charge. Assuming the corresponding $\mu_A$ is the same order
as the one from topological fluctuation, we can estimate the current as
\be
\vj_A=-D \bm {\nabla} n_A\sim D\chi\frac{\mu_A}{L} \sim T\frac{\mu_A}{L},
\ee
where $L$ is mean free path of fermions and we have taken $D\sim 1/T$ and $\c\sim T^2$. Comparing with Eq.~\eqref{eq:jA_estimate}, we conclude
if the ``$\theta$ domain'' parameter $\t_\sph/L_\theta$ is larger than $T/L$,
the current \eqref{eq:j5new} would dominate over axial current
generated by thermal diffusion.

To sum up, if the condition $\tau_{\rm sph}/L_{\theta}\gtrsim 1,\;\tau_{\rm sph}/L_{\theta}\gtrsim T/L$ is
achieved heavy-ion collisions,
the new current \eqref{eq:j5new} proposed in this paper would become
phenomenologically important.

\sect{Acknowledgments}
The authors would like to thank U.~Gursoy, C.~Hoyos, D.~Kharzeev, E.~Kiritsis, K.~Landsteiner, L.~McLerran, G.~Moore, R.~Pisarski, E.~Shuryak, H-U.~Yee and I.~Zahed for useful
discussions and the Simons Center for Geometry and Physics for
hospitality where part of this work has been done. 
I.I. would also like to thank the Mainz Institute for theoretical
Physics for the hospitality and partial support during the last stage
of this work. 
This work is supported in part by the DOE grant No. DE-FG-88ER40388
(I.I.)
and in part by DOE grant No. DE-SC0012704 (Y.Y.) .
S.L. is supported by RIKEN Foreign Postdoctoral Researcher Program.

\bibliography{Q5ref}

\begin{thebibliography}{38}%
\makeatletter
\providecommand \@ifxundefined [1]{%
 \@ifx{#1\undefined}
}%
\providecommand \@ifnum [1]{%
 \ifnum #1\expandafter \@firstoftwo
 \else \expandafter \@secondoftwo
 \fi
}%
\providecommand \@ifx [1]{%
 \ifx #1\expandafter \@firstoftwo
 \else \expandafter \@secondoftwo
 \fi
}%
\providecommand \natexlab [1]{#1}%
\providecommand \enquote  [1]{``#1''}%
\providecommand \bibnamefont  [1]{#1}%
\providecommand \bibfnamefont [1]{#1}%
\providecommand \citenamefont [1]{#1}%
\providecommand \href@noop [0]{\@secondoftwo}%
\providecommand \href [0]{\begingroup \@sanitize@url \@href}%
\providecommand \@href[1]{\@@startlink{#1}\@@href}%
\providecommand \@@href[1]{\endgroup#1\@@endlink}%
\providecommand \@sanitize@url [0]{\catcode `\\12\catcode `\$12\catcode
  `\&12\catcode `\#12\catcode `\^12\catcode `\_12\catcode `\%12\relax}%
\providecommand \@@startlink[1]{}%
\providecommand \@@endlink[0]{}%
\providecommand \url  [0]{\begingroup\@sanitize@url \@url }%
\providecommand \@url [1]{\endgroup\@href {#1}{\urlprefix }}%
\providecommand \urlprefix  [0]{URL }%
\providecommand \Eprint [0]{\href }%
\providecommand \doibase [0]{http://dx.doi.org/}%
\providecommand \selectlanguage [0]{\@gobble}%
\providecommand \bibinfo  [0]{\@secondoftwo}%
\providecommand \bibfield  [0]{\@secondoftwo}%
\providecommand \translation [1]{[#1]}%
\providecommand \BibitemOpen [0]{}%
\providecommand \bibitemStop [0]{}%
\providecommand \bibitemNoStop [0]{.\EOS\space}%
\providecommand \EOS [0]{\spacefactor3000\relax}%
\providecommand \BibitemShut  [1]{\csname bibitem#1\endcsname}%
\let\auto@bib@innerbib\@empty
\bibitem [{\citenamefont {Manton}(1983)}]{Manton:1983nd}%
  \BibitemOpen
  \bibfield  {author} {\bibinfo {author} {\bibfnamefont {N.}~\bibnamefont
  {Manton}},\ }\href {\doibase 10.1103/PhysRevD.28.2019} {\bibfield  {journal}
  {\bibinfo  {journal} {Phys.Rev.}\ }\textbf {\bibinfo {volume} {D28}},\
  \bibinfo {pages} {2019} (\bibinfo {year} {1983})}\BibitemShut {NoStop}%
\bibitem [{\citenamefont {Klinkhamer}\ and\ \citenamefont
  {Manton}(1984)}]{Klinkhamer:1984di}%
  \BibitemOpen
  \bibfield  {author} {\bibinfo {author} {\bibfnamefont {F.~R.}\ \bibnamefont
  {Klinkhamer}}\ and\ \bibinfo {author} {\bibfnamefont {N.}~\bibnamefont
  {Manton}},\ }\href {\doibase 10.1103/PhysRevD.30.2212} {\bibfield  {journal}
  {\bibinfo  {journal} {Phys.Rev.}\ }\textbf {\bibinfo {volume} {D30}},\
  \bibinfo {pages} {2212} (\bibinfo {year} {1984})}\BibitemShut {NoStop}%
\bibitem [{\citenamefont {Kuzmin}\ \emph {et~al.}(1985)\citenamefont {Kuzmin},
  \citenamefont {Rubakov},\ and\ \citenamefont {Shaposhnikov}}]{Kuzmin:1985mm}%
  \BibitemOpen
  \bibfield  {author} {\bibinfo {author} {\bibfnamefont {V.}~\bibnamefont
  {Kuzmin}}, \bibinfo {author} {\bibfnamefont {V.}~\bibnamefont {Rubakov}}, \
  and\ \bibinfo {author} {\bibfnamefont {M.}~\bibnamefont {Shaposhnikov}},\
  }\href {\doibase 10.1016/0370-2693(85)91028-7} {\bibfield  {journal}
  {\bibinfo  {journal} {Phys.Lett.}\ }\textbf {\bibinfo {volume} {B155}},\
  \bibinfo {pages} {36} (\bibinfo {year} {1985})}\BibitemShut {NoStop}%
\bibitem [{\citenamefont {Shaposhnikov}(1987)}]{Shaposhnikov:1987tw}%
  \BibitemOpen
  \bibfield  {author} {\bibinfo {author} {\bibfnamefont {M.}~\bibnamefont
  {Shaposhnikov}},\ }\href {\doibase 10.1016/0550-3213(87)90127-1} {\bibfield
  {journal} {\bibinfo  {journal} {Nucl.Phys.}\ }\textbf {\bibinfo {volume}
  {B287}},\ \bibinfo {pages} {757} (\bibinfo {year} {1987})}\BibitemShut
  {NoStop}%
\bibitem [{\citenamefont {Arnold}\ and\ \citenamefont
  {McLerran}(1987)}]{Arnold:1987mh}%
  \BibitemOpen
  \bibfield  {author} {\bibinfo {author} {\bibfnamefont {P.~B.}\ \bibnamefont
  {Arnold}}\ and\ \bibinfo {author} {\bibfnamefont {L.~D.}\ \bibnamefont
  {McLerran}},\ }\href {\doibase 10.1103/PhysRevD.36.581} {\bibfield  {journal}
  {\bibinfo  {journal} {Phys.Rev.}\ }\textbf {\bibinfo {volume} {D36}},\
  \bibinfo {pages} {581} (\bibinfo {year} {1987})}\BibitemShut {NoStop}%
\bibitem [{\citenamefont {Arnold}\ and\ \citenamefont
  {McLerran}(1988)}]{Arnold:1987zg}%
  \BibitemOpen
  \bibfield  {author} {\bibinfo {author} {\bibfnamefont {P.~B.}\ \bibnamefont
  {Arnold}}\ and\ \bibinfo {author} {\bibfnamefont {L.~D.}\ \bibnamefont
  {McLerran}},\ }\href {\doibase 10.1103/PhysRevD.37.1020} {\bibfield
  {journal} {\bibinfo  {journal} {Phys.Rev.}\ }\textbf {\bibinfo {volume}
  {D37}},\ \bibinfo {pages} {1020} (\bibinfo {year} {1988})}\BibitemShut
  {NoStop}%
\bibitem [{\citenamefont {Zhitnitsky}(2012{\natexlab{a}})}]{Zhitnitsky:2011aa}%
  \BibitemOpen
  \bibfield  {author} {\bibinfo {author} {\bibfnamefont {A.~R.}\ \bibnamefont
  {Zhitnitsky}},\ }\href {\doibase 10.1103/PhysRevD.86.045026} {\bibfield
  {journal} {\bibinfo  {journal} {Phys.Rev.}\ }\textbf {\bibinfo {volume}
  {D86}},\ \bibinfo {pages} {045026} (\bibinfo {year} {2012}{\natexlab{a}})},\
  \Eprint {http://arxiv.org/abs/1112.3365} {arXiv:1112.3365 [hep-ph]}
  \BibitemShut {NoStop}%
\bibitem [{\citenamefont {Zhitnitsky}(2012{\natexlab{b}})}]{Zhitnitsky:2012im}%
  \BibitemOpen
  \bibfield  {author} {\bibinfo {author} {\bibfnamefont {A.~R.}\ \bibnamefont
  {Zhitnitsky}},\ }\href {\doibase 10.1016/j.nuclphysa.2012.05.003} {\bibfield
  {journal} {\bibinfo  {journal} {Nucl.Phys.}\ }\textbf {\bibinfo {volume}
  {A886}},\ \bibinfo {pages} {17} (\bibinfo {year} {2012}{\natexlab{b}})},\
  \Eprint {http://arxiv.org/abs/1201.2665} {arXiv:1201.2665 [hep-ph]}
  \BibitemShut {NoStop}%
\bibitem [{\citenamefont {Zhitnitsky}(2013)}]{Zhitnitsky:2013hs}%
  \BibitemOpen
  \bibfield  {author} {\bibinfo {author} {\bibfnamefont {A.~R.}\ \bibnamefont
  {Zhitnitsky}},\ }\href {\doibase 10.1016/j.aop.2013.05.020} {\bibfield
  {journal} {\bibinfo  {journal} {Annals Phys.}\ }\textbf {\bibinfo {volume}
  {336}},\ \bibinfo {pages} {462} (\bibinfo {year} {2013})},\ \Eprint
  {http://arxiv.org/abs/1301.7072} {arXiv:1301.7072 [hep-ph]} \BibitemShut
  {NoStop}%
\bibitem [{\citenamefont {Kharzeev}\ and\ \citenamefont
  {Zhitnitsky}(2007)}]{Kharzeev:2007tn}%
  \BibitemOpen
  \bibfield  {author} {\bibinfo {author} {\bibfnamefont {D.}~\bibnamefont
  {Kharzeev}}\ and\ \bibinfo {author} {\bibfnamefont {A.}~\bibnamefont
  {Zhitnitsky}},\ }\href {\doibase 10.1016/j.nuclphysa.2007.10.001} {\bibfield
  {journal} {\bibinfo  {journal} {Nucl.Phys.}\ }\textbf {\bibinfo {volume}
  {A797}},\ \bibinfo {pages} {67} (\bibinfo {year} {2007})},\ \Eprint
  {http://arxiv.org/abs/0706.1026} {arXiv:0706.1026 [hep-ph]} \BibitemShut
  {NoStop}%
\bibitem [{\citenamefont {Liao}(2014)}]{Liao:2014ava}%
  \BibitemOpen
  \bibfield  {author} {\bibinfo {author} {\bibfnamefont {J.}~\bibnamefont
  {Liao}},\ }\href@noop {} {\  (\bibinfo {year} {2014})},\ \Eprint
  {http://arxiv.org/abs/1401.2500} {arXiv:1401.2500 [hep-ph]} \BibitemShut
  {NoStop}%
\bibitem [{\citenamefont {Chao}\ \emph {et~al.}(2013)\citenamefont {Chao},
  \citenamefont {Chu},\ and\ \citenamefont {Huang}}]{Chao:2013qpa}%
  \BibitemOpen
  \bibfield  {author} {\bibinfo {author} {\bibfnamefont {J.}~\bibnamefont
  {Chao}}, \bibinfo {author} {\bibfnamefont {P.}~\bibnamefont {Chu}}, \ and\
  \bibinfo {author} {\bibfnamefont {M.}~\bibnamefont {Huang}},\ }\href
  {\doibase 10.1103/PhysRevD.88.054009} {\bibfield  {journal} {\bibinfo
  {journal} {Phys.Rev.}\ }\textbf {\bibinfo {volume} {D88}},\ \bibinfo {pages}
  {054009} (\bibinfo {year} {2013})},\ \Eprint {http://arxiv.org/abs/1305.1100}
  {arXiv:1305.1100 [hep-ph]} \BibitemShut {NoStop}%
\bibitem [{\citenamefont {Yu}\ \emph {et~al.}(2014)\citenamefont {Yu},
  \citenamefont {Liu},\ and\ \citenamefont {Huang}}]{Yu:2014sla}%
  \BibitemOpen
  \bibfield  {author} {\bibinfo {author} {\bibfnamefont {L.}~\bibnamefont
  {Yu}}, \bibinfo {author} {\bibfnamefont {H.}~\bibnamefont {Liu}}, \ and\
  \bibinfo {author} {\bibfnamefont {M.}~\bibnamefont {Huang}},\ }\href
  {\doibase 10.1103/PhysRevD.90.074009} {\bibfield  {journal} {\bibinfo
  {journal} {Phys.Rev.}\ }\textbf {\bibinfo {volume} {D90}},\ \bibinfo {pages}
  {074009} (\bibinfo {year} {2014})},\ \Eprint {http://arxiv.org/abs/1404.6969}
  {arXiv:1404.6969 [hep-ph]} \BibitemShut {NoStop}%
\bibitem [{\citenamefont {Fukushima}\ \emph {et~al.}(2008)\citenamefont
  {Fukushima}, \citenamefont {Kharzeev},\ and\ \citenamefont
  {Warringa}}]{Fukushima:2008xe}%
  \BibitemOpen
  \bibfield  {author} {\bibinfo {author} {\bibfnamefont {K.}~\bibnamefont
  {Fukushima}}, \bibinfo {author} {\bibfnamefont {D.~E.}\ \bibnamefont
  {Kharzeev}}, \ and\ \bibinfo {author} {\bibfnamefont {H.~J.}\ \bibnamefont
  {Warringa}},\ }\href {\doibase 10.1103/PhysRevD.78.074033} {\bibfield
  {journal} {\bibinfo  {journal} {Phys. Rev. D}\ }\textbf {\bibinfo {volume}
  {78}},\ \bibinfo {pages} {074033} (\bibinfo {year} {2008})},\ \Eprint
  {http://arxiv.org/abs/0808.3382} {arXiv:0808.3382 [hep-ph]} \BibitemShut
  {NoStop}%
\bibitem [{\citenamefont {Kharzeev}(2014)}]{Kharzeev:2013ffa}%
  \BibitemOpen
  \bibfield  {author} {\bibinfo {author} {\bibfnamefont {D.~E.}\ \bibnamefont
  {Kharzeev}},\ }\href {\doibase 10.1016/j.ppnp.2014.01.002} {\bibfield
  {journal} {\bibinfo  {journal} {Prog.Part.Nucl.Phys.}\ }\textbf {\bibinfo
  {volume} {75}},\ \bibinfo {pages} {133} (\bibinfo {year} {2014})},\ \Eprint
  {http://arxiv.org/abs/1312.3348} {arXiv:1312.3348 [hep-ph]} \BibitemShut
  {NoStop}%
\bibitem [{\citenamefont {Fukushima}\ \emph {et~al.}(2010)\citenamefont
  {Fukushima}, \citenamefont {Kharzeev},\ and\ \citenamefont
  {Warringa}}]{Fukushima:2010vw}%
  \BibitemOpen
  \bibfield  {author} {\bibinfo {author} {\bibfnamefont {K.}~\bibnamefont
  {Fukushima}}, \bibinfo {author} {\bibfnamefont {D.~E.}\ \bibnamefont
  {Kharzeev}}, \ and\ \bibinfo {author} {\bibfnamefont {H.~J.}\ \bibnamefont
  {Warringa}},\ }\href {\doibase 10.1103/PhysRevLett.104.212001} {\bibfield
  {journal} {\bibinfo  {journal} {Phys.Rev.Lett.}\ }\textbf {\bibinfo {volume}
  {104}},\ \bibinfo {pages} {212001} (\bibinfo {year} {2010})},\ \Eprint
  {http://arxiv.org/abs/1002.2495} {arXiv:1002.2495 [hep-ph]} \BibitemShut
  {NoStop}%
\bibitem [{\citenamefont {Kharzeev}\ \emph {et~al.}(2008)\citenamefont
  {Kharzeev}, \citenamefont {McLerran},\ and\ \citenamefont
  {Warringa}}]{Kharzeev:2007jp}%
  \BibitemOpen
  \bibfield  {author} {\bibinfo {author} {\bibfnamefont {D.~E.}\ \bibnamefont
  {Kharzeev}}, \bibinfo {author} {\bibfnamefont {L.~D.}\ \bibnamefont
  {McLerran}}, \ and\ \bibinfo {author} {\bibfnamefont {H.~J.}\ \bibnamefont
  {Warringa}},\ }\href {\doibase 10.1016/j.nuclphysa.2008.02.298} {\bibfield
  {journal} {\bibinfo  {journal} {Nucl.Phys.}\ }\textbf {\bibinfo {volume}
  {A803}},\ \bibinfo {pages} {227} (\bibinfo {year} {2008})},\ \Eprint
  {http://arxiv.org/abs/0711.0950} {arXiv:0711.0950 [hep-ph]} \BibitemShut
  {NoStop}%
\bibitem [{\citenamefont {Bonati}\ \emph {et~al.}(2014)\citenamefont {Bonati},
  \citenamefont {D'Elia}, \citenamefont {Panagopoulos},\ and\ \citenamefont
  {Vicari}}]{Bonati:2013dza}%
  \BibitemOpen
  \bibfield  {author} {\bibinfo {author} {\bibfnamefont {C.}~\bibnamefont
  {Bonati}}, \bibinfo {author} {\bibfnamefont {M.}~\bibnamefont {D'Elia}},
  \bibinfo {author} {\bibfnamefont {H.}~\bibnamefont {Panagopoulos}}, \ and\
  \bibinfo {author} {\bibfnamefont {E.}~\bibnamefont {Vicari}},\ }\href@noop {}
  {\bibfield  {journal} {\bibinfo  {journal} {PoS}\ }\textbf {\bibinfo {volume}
  {LATTICE2013}},\ \bibinfo {pages} {136} (\bibinfo {year} {2014})},\ \Eprint
  {http://arxiv.org/abs/1309.6059} {arXiv:1309.6059 [hep-lat]} \BibitemShut
  {NoStop}%
\bibitem [{\citenamefont {Bergman}\ and\ \citenamefont
  {Lifschytz}(2007)}]{Bergman:2006xn}%
  \BibitemOpen
  \bibfield  {author} {\bibinfo {author} {\bibfnamefont {O.}~\bibnamefont
  {Bergman}}\ and\ \bibinfo {author} {\bibfnamefont {G.}~\bibnamefont
  {Lifschytz}},\ }\href {\doibase 10.1088/1126-6708/2007/04/043} {\bibfield
  {journal} {\bibinfo  {journal} {JHEP}\ }\textbf {\bibinfo {volume} {0704}},\
  \bibinfo {pages} {043} (\bibinfo {year} {2007})},\ \Eprint
  {http://arxiv.org/abs/hep-th/0612289} {arXiv:hep-th/0612289 [hep-th]}
  \BibitemShut {NoStop}%
\bibitem [{\citenamefont {Vicari}\ and\ \citenamefont
  {Panagopoulos}(2009)}]{Vicari:2008jw}%
  \BibitemOpen
  \bibfield  {author} {\bibinfo {author} {\bibfnamefont {E.}~\bibnamefont
  {Vicari}}\ and\ \bibinfo {author} {\bibfnamefont {H.}~\bibnamefont
  {Panagopoulos}},\ }\href {\doibase 10.1016/j.physrep.2008.10.001} {\bibfield
  {journal} {\bibinfo  {journal} {Phys.Rept.}\ }\textbf {\bibinfo {volume}
  {470}},\ \bibinfo {pages} {93} (\bibinfo {year} {2009})},\ \Eprint
  {http://arxiv.org/abs/0803.1593} {arXiv:0803.1593 [hep-th]} \BibitemShut
  {NoStop}%
\bibitem [{\citenamefont {Rubakov}\ and\ \citenamefont
  {Shaposhnikov}(1996)}]{Rubakov:1996vz}%
  \BibitemOpen
  \bibfield  {author} {\bibinfo {author} {\bibfnamefont {V.}~\bibnamefont
  {Rubakov}}\ and\ \bibinfo {author} {\bibfnamefont {M.}~\bibnamefont
  {Shaposhnikov}},\ }\href {\doibase 10.1070/PU1996v039n05ABEH000145}
  {\bibfield  {journal} {\bibinfo  {journal} {Usp.Fiz.Nauk}\ }\textbf {\bibinfo
  {volume} {166}},\ \bibinfo {pages} {493} (\bibinfo {year} {1996})},\ \Eprint
  {http://arxiv.org/abs/hep-ph/9603208} {arXiv:hep-ph/9603208 [hep-ph]}
  \BibitemShut {NoStop}%
\bibitem [{\citenamefont {Iatrakis}\ \emph {et~al.}(2015)\citenamefont
  {Iatrakis}, \citenamefont {Lin},\ and\ \citenamefont
  {Yin}}]{Iatrakis:2015fma}%
  \BibitemOpen
  \bibfield  {author} {\bibinfo {author} {\bibfnamefont {I.}~\bibnamefont
  {Iatrakis}}, \bibinfo {author} {\bibfnamefont {S.}~\bibnamefont {Lin}}, \
  and\ \bibinfo {author} {\bibfnamefont {Y.}~\bibnamefont {Yin}},\ }\href@noop
  {} {\  (\bibinfo {year} {2015})},\ \Eprint {http://arxiv.org/abs/1506.01384}
  {arXiv:1506.01384 [hep-th]} \BibitemShut {NoStop}%
\bibitem [{Note1()}]{Note1}%
  \BibitemOpen
  \bibinfo {note} {We stress that our $\theta $ field originates from
  topological fluctuation of gluons. It should not be confused with $\theta $
  field from fermionic quasi-zero mode in \cite
  {Kalaydzhyan:2012ut}}\BibitemShut {NoStop}%
\bibitem [{\citenamefont {Sakai}\ and\ \citenamefont
  {Sugimoto}(2005{\natexlab{a}})}]{Sakai:2004cn}%
  \BibitemOpen
  \bibfield  {author} {\bibinfo {author} {\bibfnamefont {T.}~\bibnamefont
  {Sakai}}\ and\ \bibinfo {author} {\bibfnamefont {S.}~\bibnamefont
  {Sugimoto}},\ }\href {\doibase 10.1143/PTP.113.843} {\bibfield  {journal}
  {\bibinfo  {journal} {Prog.Theor.Phys.}\ }\textbf {\bibinfo {volume} {113}},\
  \bibinfo {pages} {843} (\bibinfo {year} {2005}{\natexlab{a}})},\ \Eprint
  {http://arxiv.org/abs/hep-th/0412141} {arXiv:hep-th/0412141 [hep-th]}
  \BibitemShut {NoStop}%
\bibitem [{\citenamefont {Sakai}\ and\ \citenamefont
  {Sugimoto}(2005{\natexlab{b}})}]{Sakai:2005yt}%
  \BibitemOpen
  \bibfield  {author} {\bibinfo {author} {\bibfnamefont {T.}~\bibnamefont
  {Sakai}}\ and\ \bibinfo {author} {\bibfnamefont {S.}~\bibnamefont
  {Sugimoto}},\ }\href {\doibase 10.1143/PTP.114.1083} {\bibfield  {journal}
  {\bibinfo  {journal} {Prog.Theor.Phys.}\ }\textbf {\bibinfo {volume} {114}},\
  \bibinfo {pages} {1083} (\bibinfo {year} {2005}{\natexlab{b}})},\ \Eprint
  {http://arxiv.org/abs/hep-th/0507073} {arXiv:hep-th/0507073 [hep-th]}
  \BibitemShut {NoStop}%
\bibitem [{\citenamefont {Witten}(1998)}]{Witten:1998zw}%
  \BibitemOpen
  \bibfield  {author} {\bibinfo {author} {\bibfnamefont {E.}~\bibnamefont
  {Witten}},\ }\href@noop {} {\bibfield  {journal} {\bibinfo  {journal}
  {Adv.Theor.Math.Phys.}\ }\textbf {\bibinfo {volume} {2}},\ \bibinfo {pages}
  {505} (\bibinfo {year} {1998})},\ \Eprint
  {http://arxiv.org/abs/hep-th/9803131} {arXiv:hep-th/9803131 [hep-th]}
  \BibitemShut {NoStop}%
\bibitem [{\citenamefont {Aharony}\ \emph {et~al.}(2007)\citenamefont
  {Aharony}, \citenamefont {Sonnenschein},\ and\ \citenamefont
  {Yankielowicz}}]{Aharony:2006da}%
  \BibitemOpen
  \bibfield  {author} {\bibinfo {author} {\bibfnamefont {O.}~\bibnamefont
  {Aharony}}, \bibinfo {author} {\bibfnamefont {J.}~\bibnamefont
  {Sonnenschein}}, \ and\ \bibinfo {author} {\bibfnamefont {S.}~\bibnamefont
  {Yankielowicz}},\ }\href {\doibase 10.1016/j.aop.2006.11.002} {\bibfield
  {journal} {\bibinfo  {journal} {Annals Phys.}\ }\textbf {\bibinfo {volume}
  {322}},\ \bibinfo {pages} {1420} (\bibinfo {year} {2007})},\ \Eprint
  {http://arxiv.org/abs/hep-th/0604161} {arXiv:hep-th/0604161 [hep-th]}
  \BibitemShut {NoStop}%
\bibitem [{Note2()}]{Note2}%
  \BibitemOpen
  \bibinfo {note} {The Levi-Civita symbol is fixed by $\epsilon
  ^{txu}=1$.}\BibitemShut {Stop}%
\bibitem [{\citenamefont {Kanitscheider}\ \emph {et~al.}(2008)\citenamefont
  {Kanitscheider}, \citenamefont {Skenderis},\ and\ \citenamefont
  {Taylor}}]{Kanitscheider:2008kd}%
  \BibitemOpen
  \bibfield  {author} {\bibinfo {author} {\bibfnamefont {I.}~\bibnamefont
  {Kanitscheider}}, \bibinfo {author} {\bibfnamefont {K.}~\bibnamefont
  {Skenderis}}, \ and\ \bibinfo {author} {\bibfnamefont {M.}~\bibnamefont
  {Taylor}},\ }\href {\doibase 10.1088/1126-6708/2008/09/094} {\bibfield
  {journal} {\bibinfo  {journal} {JHEP}\ }\textbf {\bibinfo {volume} {0809}},\
  \bibinfo {pages} {094} (\bibinfo {year} {2008})},\ \Eprint
  {http://arxiv.org/abs/0807.3324} {arXiv:0807.3324 [hep-th]} \BibitemShut
  {NoStop}%
\bibitem [{\citenamefont {Kovtun}\ and\ \citenamefont
  {Starinets}(2005)}]{Kovtun:2005ev}%
  \BibitemOpen
  \bibfield  {author} {\bibinfo {author} {\bibfnamefont {P.~K.}\ \bibnamefont
  {Kovtun}}\ and\ \bibinfo {author} {\bibfnamefont {A.~O.}\ \bibnamefont
  {Starinets}},\ }\href {\doibase 10.1103/PhysRevD.72.086009} {\bibfield
  {journal} {\bibinfo  {journal} {Phys.Rev.}\ }\textbf {\bibinfo {volume}
  {D72}},\ \bibinfo {pages} {086009} (\bibinfo {year} {2005})},\ \Eprint
  {http://arxiv.org/abs/hep-th/0506184} {arXiv:hep-th/0506184 [hep-th]}
  \BibitemShut {NoStop}%
\bibitem [{\citenamefont {Iqbal}\ and\ \citenamefont
  {Liu}(2009)}]{Iqbal:2008by}%
  \BibitemOpen
  \bibfield  {author} {\bibinfo {author} {\bibfnamefont {N.}~\bibnamefont
  {Iqbal}}\ and\ \bibinfo {author} {\bibfnamefont {H.}~\bibnamefont {Liu}},\
  }\href {\doibase 10.1103/PhysRevD.79.025023} {\bibfield  {journal} {\bibinfo
  {journal} {Phys.Rev.}\ }\textbf {\bibinfo {volume} {D79}},\ \bibinfo {pages}
  {025023} (\bibinfo {year} {2009})},\ \Eprint {http://arxiv.org/abs/0809.3808}
  {arXiv:0809.3808 [hep-th]} \BibitemShut {NoStop}%
\bibitem [{Note3()}]{Note3}%
  \BibitemOpen
  \bibinfo {note} {The value of CS diffusion rate $\Gamma _\protect \text {CS}$
  has already been calculated in Ref.~\cite {Craps:2012hd}. Our value is four
  times the value computed in Ref.~\cite {Craps:2012hd} due to a different
  normalization.}\BibitemShut {Stop}%
\bibitem [{\citenamefont {Yee}(2009)}]{Yee:2009vw}%
  \BibitemOpen
  \bibfield  {author} {\bibinfo {author} {\bibfnamefont {H.-U.}\ \bibnamefont
  {Yee}},\ }\href {\doibase 10.1088/1126-6708/2009/11/085} {\bibfield
  {journal} {\bibinfo  {journal} {JHEP}\ }\textbf {\bibinfo {volume} {0911}},\
  \bibinfo {pages} {085} (\bibinfo {year} {2009})},\ \Eprint
  {http://arxiv.org/abs/0908.4189} {arXiv:0908.4189 [hep-th]} \BibitemShut
  {NoStop}%
\bibitem [{\citenamefont {Son}\ and\ \citenamefont
  {Zhitnitsky}(2004)}]{Son:2004tq}%
  \BibitemOpen
  \bibfield  {author} {\bibinfo {author} {\bibfnamefont {D.}~\bibnamefont
  {Son}}\ and\ \bibinfo {author} {\bibfnamefont {A.~R.}\ \bibnamefont
  {Zhitnitsky}},\ }\href {\doibase 10.1103/PhysRevD.70.074018} {\bibfield
  {journal} {\bibinfo  {journal} {Phys.Rev.}\ }\textbf {\bibinfo {volume}
  {D70}},\ \bibinfo {pages} {074018} (\bibinfo {year} {2004})},\ \Eprint
  {http://arxiv.org/abs/hep-ph/0405216} {arXiv:hep-ph/0405216 [hep-ph]}
  \BibitemShut {NoStop}%
\bibitem [{\citenamefont {Metlitski}\ and\ \citenamefont
  {Zhitnitsky}(2005)}]{Metlitski:2005pr}%
  \BibitemOpen
  \bibfield  {author} {\bibinfo {author} {\bibfnamefont {M.~A.}\ \bibnamefont
  {Metlitski}}\ and\ \bibinfo {author} {\bibfnamefont {A.~R.}\ \bibnamefont
  {Zhitnitsky}},\ }\href {\doibase 10.1103/PhysRevD.72.045011} {\bibfield
  {journal} {\bibinfo  {journal} {Phys.Rev.}\ }\textbf {\bibinfo {volume}
  {D72}},\ \bibinfo {pages} {045011} (\bibinfo {year} {2005})},\ \Eprint
  {http://arxiv.org/abs/hep-ph/0505072} {arXiv:hep-ph/0505072 [hep-ph]}
  \BibitemShut {NoStop}%
\bibitem [{\citenamefont {Huang}\ and\ \citenamefont
  {Liao}(2013)}]{Huang:2013iia}%
  \BibitemOpen
  \bibfield  {author} {\bibinfo {author} {\bibfnamefont {X.-G.}\ \bibnamefont
  {Huang}}\ and\ \bibinfo {author} {\bibfnamefont {J.}~\bibnamefont {Liao}},\
  }\href {\doibase 10.1103/PhysRevLett.110.232302} {\bibfield  {journal}
  {\bibinfo  {journal} {Phys.Rev.Lett.}\ }\textbf {\bibinfo {volume} {110}},\
  \bibinfo {pages} {232302} (\bibinfo {year} {2013})},\ \Eprint
  {http://arxiv.org/abs/1303.7192} {arXiv:1303.7192 [nucl-th]} \BibitemShut
  {NoStop}%
\bibitem [{\citenamefont {Kalaydzhyan}(2013)}]{Kalaydzhyan:2012ut}%
  \BibitemOpen
  \bibfield  {author} {\bibinfo {author} {\bibfnamefont {T.}~\bibnamefont
  {Kalaydzhyan}},\ }\href {\doibase 10.1016/j.nuclphysa.2013.06.009} {\bibfield
   {journal} {\bibinfo  {journal} {Nucl.Phys.}\ }\textbf {\bibinfo {volume}
  {A913}},\ \bibinfo {pages} {243} (\bibinfo {year} {2013})},\ \Eprint
  {http://arxiv.org/abs/1208.0012} {arXiv:1208.0012 [hep-ph]} \BibitemShut
  {NoStop}%
\bibitem [{\citenamefont {Craps}\ \emph {et~al.}(2012)\citenamefont {Craps},
  \citenamefont {Hoyos}, \citenamefont {Surowka},\ and\ \citenamefont
  {Taels}}]{Craps:2012hd}%
  \BibitemOpen
  \bibfield  {author} {\bibinfo {author} {\bibfnamefont {B.}~\bibnamefont
  {Craps}}, \bibinfo {author} {\bibfnamefont {C.}~\bibnamefont {Hoyos}},
  \bibinfo {author} {\bibfnamefont {P.}~\bibnamefont {Surowka}}, \ and\
  \bibinfo {author} {\bibfnamefont {P.}~\bibnamefont {Taels}},\ }\href
  {\doibase 10.1007/JHEP11(2012)109, 10.1007/JHEP02(2013)087} {\bibfield
  {journal} {\bibinfo  {journal} {JHEP}\ }\textbf {\bibinfo {volume} {1211}},\
  \bibinfo {pages} {109} (\bibinfo {year} {2012})},\ \Eprint
  {http://arxiv.org/abs/1209.2532} {arXiv:1209.2532 [hep-th]} \BibitemShut
  {NoStop}%
\end{thebibliography}%

\end{document}